\documentclass[twocolumn, prb, showpacs,amsmath,amssymb, hidelinks]{revtex4}
\usepackage[dvips]{graphicx}
\usepackage{pslatex}
\usepackage{amsmath}    % need for subequations
\usepackage{MnSymbol} % simboli delle caption
\usepackage{graphicx}    % need for figures
\usepackage{verbatim}    % useful for program listings
\usepackage{color}           % use if color is used in text
\usepackage{subfigure}   % use for side-by-side figures
\usepackage{hyperref}    % use for hypertext links, including those to external documents and URLs

\begin{document}

%Title of paper
\title{Strain sensitivity and superconducting properties of Nb$_3$Sn from first principles calculations}

\author{G. De Marzi}
\email[]{demarzi@enea.it}
\author{L. Morici}
\author{L. Muzzi}
\author{A. della Corte}
\affiliation{EURATOM-ENEA Association on Fusion, Via Enrico Fermi 45, 00044 Frascati RM, Italy}
\author{M. Buongiorno Nardelli}
\affiliation{Department of Physics, University of North Texas, 1155 Union Circle 311427, Denton, Texas 76203-5017, USA}
\address{CSMD, Oak Ridge National Laboratory, Oak Ridge, TN 37831}
\date{\today}

\begin{abstract}

Using calculations from first principles based on density functional theory we have studied the strain sensitivity of the high-field superconducting magnet A15 Nb$_3$Sn. 
The Nb$_3$Sn lattice cell was deformed in the same way as observed experimentally on multi-filamentary, technological wires subject to loads applied along their axes. The phonon dispersion curves and electronic band structures along different high-symmetry directions in the Brillouin zone were calculated, at different levels of applied strain, $\epsilon$, both on the compressive and the tensile side. Starting from the calculated averaged phonon frequencies and electron-phonon coupling, the superconducting characteristic critical temperature of the material, $T_c$, has been calculated by means of the Allen-Dynes modification of the McMillan formula. As a result, the characteristic bell-shaped $T_c$ vs. $\epsilon$ curve, with a maximum at zero intrinsic strain, and with a slight asymmetry between the tensile and compressive sides, has been obtained. These first-principle calculations thus show that the strain sensitivity of Nb$_3$Sn has a microscopic and intrinsic origin, originating from shifts in the Nb$_3$Sn critical surface. In addition, our computations show that variations of superconducting properties of this compound are correlated to stress-induced changes in both the phononic and electronic properties.
Finally, the strain function describing the strain sensitivity of Nb$_3$Sn has been extracted from the computed $T_c(\epsilon)$ curve, and compared to experimental data from multi-filamentary, composite wires. Both curves show the expected bell-shaped behavior, but the strain sensitivity of the wire is enhanced with respect to the theoretical predictions of the bulk, perfectly binary and stoichiometric Nb$_3$Sn. Understanding the origin of this difference might open potential pathways towards the improvement of the strain tolerance in such systems.

\end{abstract}

\pacs{74.70.Ad, 74.25.Kc, 74.25.Jb, 68.35.Gy, 63.20.kd, 63.20.dk, 71.15.Mb}
%\keywords{A15 superconductors, electron-phonon interaction, strain sensitivity, phonon dispersion curves, electronic band structures}

\maketitle

\section{INTRODUCTION}

The A15 phase Nb$_3$Sn compound \cite{Matthias53} is currently being used in a variety of large-scale scientific projects employing high-field superconducting magnets (above 10 T) \cite{miyazaki03}, including  ITER (the International Thermonuclear Experimental Reactor) \cite{vostner06, mitchell12, devred12}, the 1 GHz NMR project \cite{wada02}, and the CERN LHC Luminosity Upgrade \cite{bottura12}.
In these high-field magnets, the mechanical loads during cooldown (due to different thermal contractions) and operation (due to Lorentz forces) can be very large, and since the superconducting properties of Nb$_3$Sn strongly depend on strain \cite{rupp77, ekin79, tenhaken94, ekin87}, an overall performance degradation can take place.
Therefore, for a magnet's sound design it is of fundamental importance to have knowledge of the behavior of the superconducting parameters (namely the critical temperature,T$_c$, the upper critical field, B$_{c2}$, and the critical current, I$_c$) as a function of strain, $\varepsilon$.

\begin{figure}[hb]
\begin{center}
\scalebox{0.32}{\includegraphics{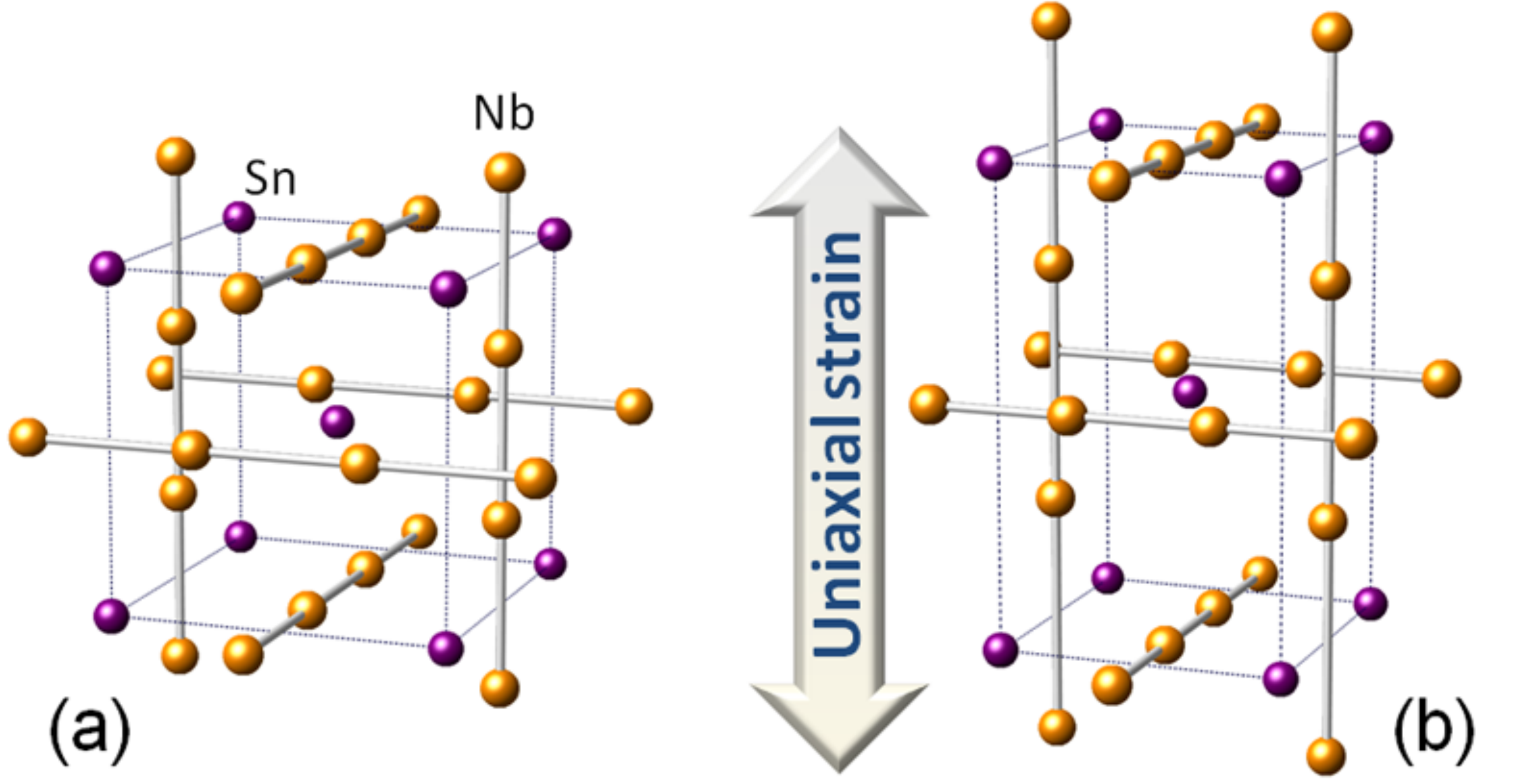}}
\caption{\label{fig:1}The arrangement of atoms in the cubic (a) and tetragonal (b) phase of A15 Nb$_3$Sn. The Nb atoms in the 2e sites form chains along the [001] direction, whereas those in the 4k sites are in chains along the [100] and [010] directions. To make the distortion clear, in (b), the cell is stretched to the abnormal value of $\epsilon$ = 40\%, whereas the Poisson's ratio $\nu$ is set to 0.4.}
\end{center}
\end{figure}

Of particular interest is the uniaxial stress (either in tension or compression), acting along the axial direction of the composite, multi-filamentary wires used in such systems: in a Cable-in-Conduit Conductor (CICC) \cite{hoenig79, bruzzone06, spadoni94}, for example, the Nb$_3$Sn wires are inserted into stainless steel conduits, and compressive stresses due to the different thermal contraction coefficients of the different materials become important \cite{nijhuis06, mitchell05}. Trasverse load components might also be important \cite{ekin87, mondonico12}, but we will focus here only on the uniaxial ones. 

\begin{figure*}[ht]
\begin{center}
\scalebox{0.55}{\includegraphics{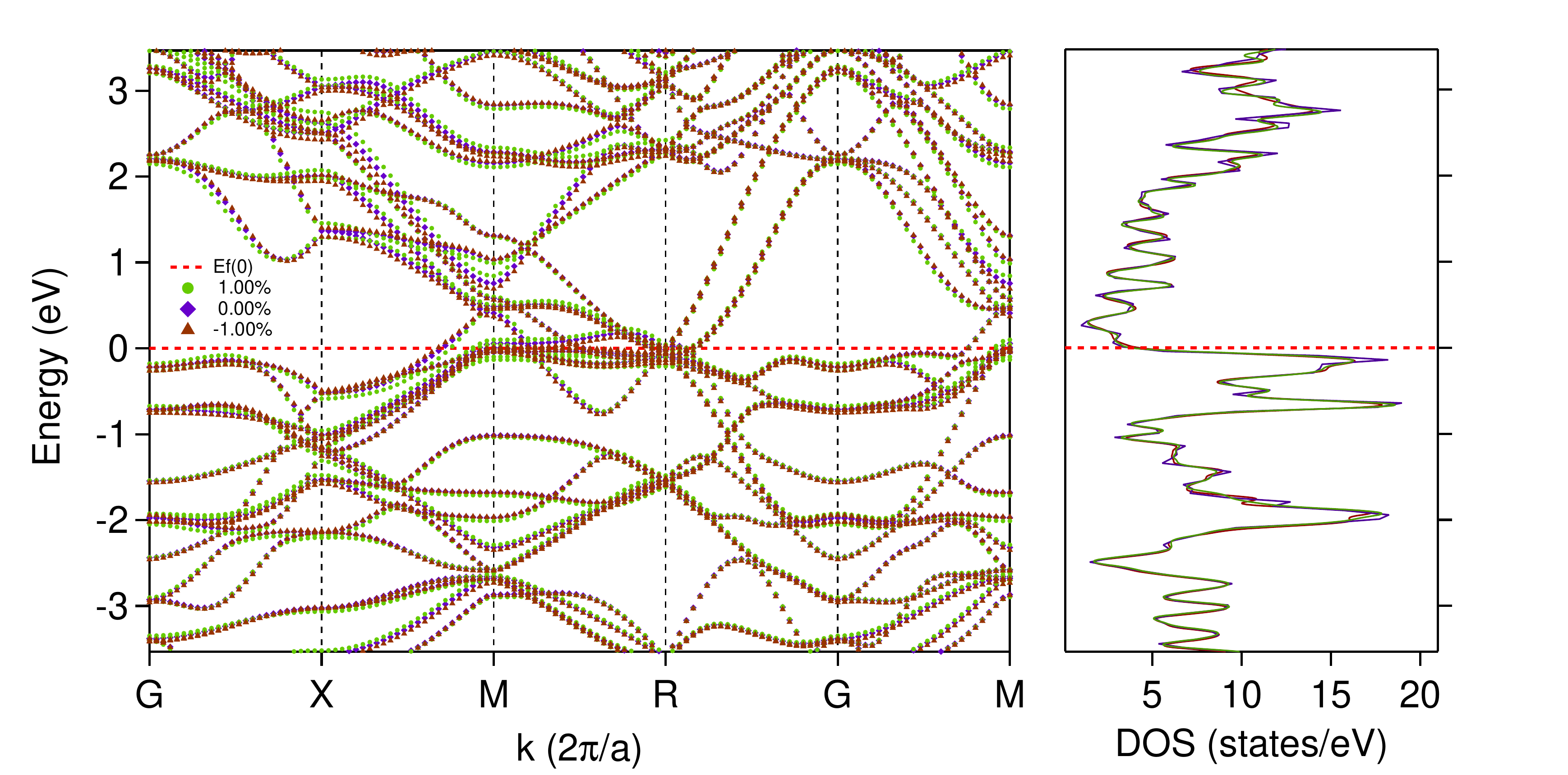}}
\caption{\label{fig:2}The electronic band structure and density of states for Nb$_3$Sn calculated for three representative strain states: +1.0\% ($\bullet$), zero applied strain ($\filleddiamond$), and -1.0\% ($\filledtriangleup$). The Fermi level is set to 0 eV.}
\end{center}
\end{figure*}

Within the framework of the Unified Scaling Law \cite{ekin80}, many authors \cite{ekin80, taylor05, oh06, godeke06, arbelaez09, markiewicz06} have proposed modified scaling equations which take into account the uniaxial strain dependence through the so-called strain function, s($\varepsilon$), but few attempts have been made \cite{oh06, taylor05, markiewicz04} on obtaining a scaling law based on microscopic parameters. For this purpose, a very first step would be to accurately determine the electronic band structures, the phonon dispersion curves and the electron-phonon coupling terms, and study their evolution as a function of applied strain. To this aim, the knowledge of the Nb$_3$Sn lattice cell deformation when a multifilamentary wire is subject to different stress components is of basic importance. Recent high resolution X-ray diffraction experiments on mechanically loaded samples \cite{muzzi12} have shown in detail how the Nb$_3$Sn lattice cell deforms in the axial and the transverse directions; in particular, it was observed that the stress is completely transferred from the macroscopic level to the individual grains within the composite structure, so that a macroscopic uniaxial load directly corresponds to a stretching of the Nb$_3$Sn lattice cell along the same direction, with the cell contracting in the transverse direction of an amount corresponding to a Poisson's ratio $\nu$ equal to 0.38.

The structural and electronic properties of Nb$_3$Sn have been theoretically studied by several groups \cite{klein01, sadigh98, lu97, mattheiss82, klein78, mattheiss75}, whereas the full phonon dispersion relations have been calculated by means of a tight-binding method \cite{weber84, weber84book} and - more recently - by an \emph{ab initio} pseudopotential approach \cite{tutuncu06}. In particular, calculations by T\"ut\"unc\"u \emph{et al.} give evidence of a strong interaction between the electronic states near the Fermi level and several phonon modes (longitudinal acoustic phonons and a group of optical phonon modes with average frequency of 4.5 THz) along the [111] direction.

However, to the best of our knowledge no systematic \emph{ab initio} investigations have been made on studying the evolution of the band structure, phonon dispersion curves and superconducting parameters (electron-phonon mass enhancement parameter, $\lambda$, and T$_c$) as a function of an applied uniaxial strain. In the present work, this issue has been addressed by employing the plane-wave pseudopotential method, the density-functional theory, and a linear-response technique \cite{baroni87, baroni01}, and by using the results by T\"ut\"unc\"u \emph{et al.} for the undistorted cell as a starting baseline for our calculations.

\section{DETAILS OF CALCULATIONS AND COMPUTATIONAL METHOD}

We used density functional theory and density-functional perturbation theory \cite{baroni01} as implemented in the  {\sc Quantum-ESPRESSO}  software distribution \cite{QE}, within the local-density approximation \cite{troullier91}, a plane-wave expansion up to 40 Ry for the kinetic energy cutoff and ultrasoft pseudopotentials for Nb and Sn \cite{pseudo}.
The Brillouin zone has been sampled on a 8$\times$8$\times$8 Monkhorst-Pack (MP) mesh, corresponding to 126 special \textbf{k}-points within the irreducible part of the Brillouin zone (IBZ). We also have checked more dense grids (up to 16$\times$16$\times$16) but the results did not change considerably.

Lattice dynamical calculations have been performed within the framework of the self-consistent density functional perturbation theory (DFPT) \cite{baroni01}, in which the dynamical matrices are calculated by sampling the IBZ with 8 independent $\textbf{q}$-points in the tetragonal phase. Dynamical matrices at any wave vectors can be Fourier deconvolved on this mesh, and the phonon dispersion curves along arbitrary symmetry directions can be easily obtained. In order to check the accuracy of the Fourier interpolation, we compared the results of this procedure with direct calculations on selected $\textbf{q}$-points not present in the grid. 

\begin{figure*}[t]
\begin{center}
\scalebox{0.55}{\includegraphics{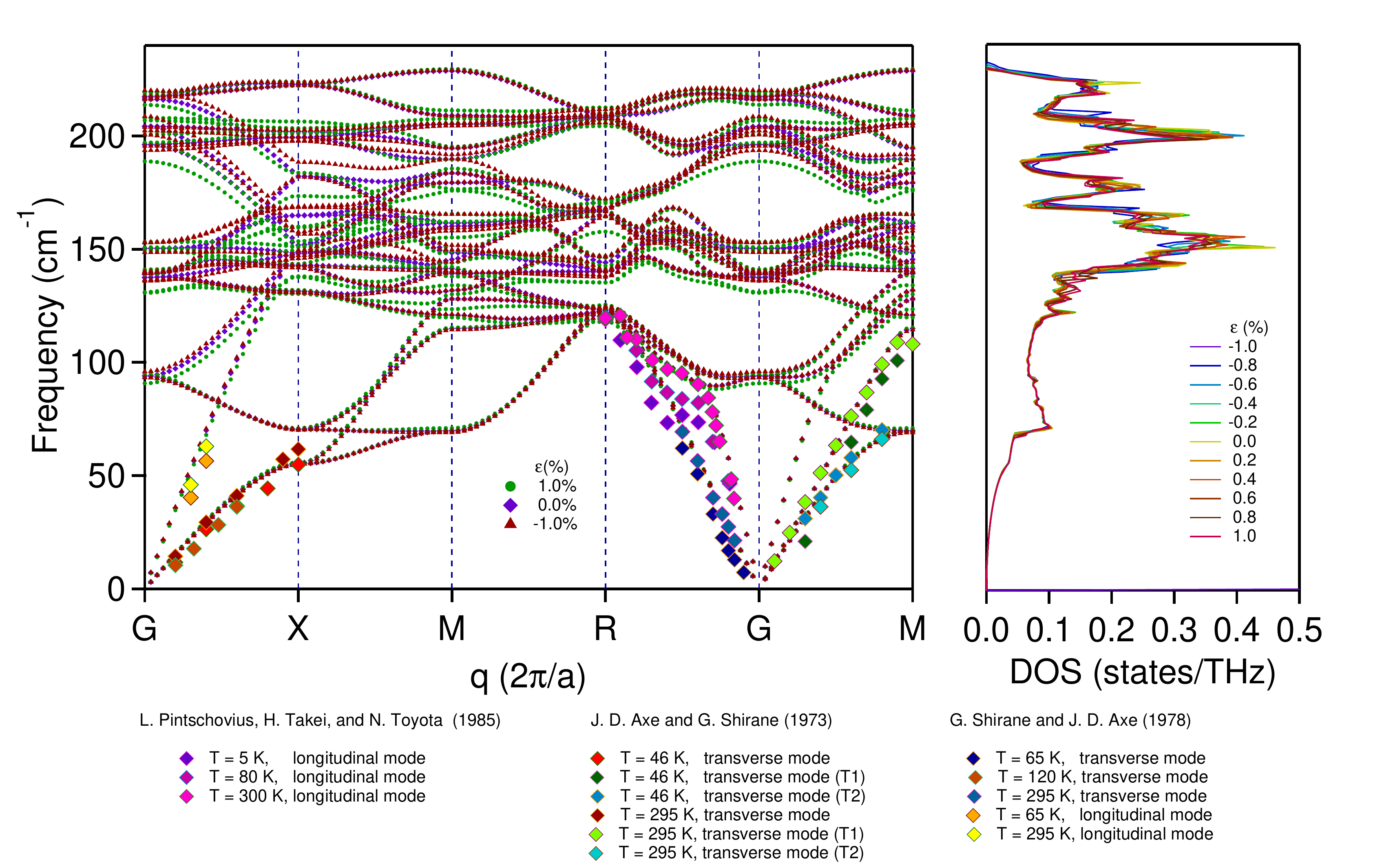}}
\caption{\label{fig:3} (a) Phonon dispersion curves of Nb$_3$Sn at three different strain states: +1.0\% ($\bullet$), zero applied strain ($\filleddiamond$), and -1.0\% ($\filledtriangleup$); (b) phonon DOS at each calculated strain. For completeness, experimental curves are also reported.}
\end{center}
\end{figure*}

A denser grid of \textbf{k}-points (24$\times$24$\times$24 MP divisions) has been used in order to determine the electron-phonon interaction parameter $\lambda$, calculated as the Brillouin-zone average of the mode-resolved coupling strengths $\lambda_{\textbf{q}j}$:

\begin{equation}
   \lambda = \sum_{\textbf{q}j} W(\textbf{q}) \lambda_{\textbf{q}j}
\end{equation}

where $\emph{j}$ indicates a phonon polarization branch, and W(\textbf{q}) are the weights associated with the phonon wavevectors \textbf{q}, normalized to 1 in the first Brillouin zone.

The lattice parameter of the cubic cell is set to 5.29 $\AA$ \cite{maier69, guritanu04}. For an uniaxial stress along the \emph{z}-direction ($\sigma_z$) and under the assumption that the system is transversally isotropic, the strain state can be expressed as:

\begin{equation}\label{stress-strain}
\begin{aligned}
      &\epsilon_x =  \epsilon_y = -\nu\frac{\sigma_z}{E}\\
     & \epsilon_z = \frac{\sigma_z}{E}
      \end{aligned}
\end{equation}

which reflects the variation of the lattice parameters of the tetragonally distorted cell.
 In Eq. \ref{stress-strain}, $\nu$ is the Poisson's ratio whereas $\emph{E}$ represents the Young's modulus. In our computations, $\nu$ has been set to the value measured in composite wires ($\nu$ = 0.4 \cite{muzzi12}), and the distortions have been calculated in the strain range $\pm$ 1.0\%, in steps of 0.2\%.

\section{RESULTS AND DISCUSSION}

The Nb$_3$Sn cubic phase belongs to the $(P\frac{4_2}{m}\bar3\frac{2}{n})$ space group and the $O_h^3$ point-group symmetries, as shown in Fig.\ref{fig:1}(a). The Sn atoms are situated on a bcc matrix whereas the faces of the cube are occupied by Nb atoms which form three sets of orthogonal chains along the principal axes. When a uniaxial strain is applied along the c-direction, the lattice is tetragonally distorted, with the Nb-chains in the [001] direction differing from those in the [100] and [010] directions, as shown in Fig. \ref{fig:1}(b). The distorted structure has a reduced symmetry $D_{4h}^9 (P\frac{4_2}{m}\frac{2}{m}\frac{2}{c})$.
 
Starting from the cubic cell, a uniaxial strain has been applied to the cell along the \emph{c}-direction, according to Eq. (\ref{stress-strain}). As a result, the deviatoric components of the strain lowered the system's point group symmetries: most of the phonon degeneracies have been removed and a changement in both the electronic and phonon dispersion bands have been induced.

\subsection{Electronic band structures and phonon dispersion curves}

\begin{figure}[t]
\begin{center}
\scalebox{0.5}{\includegraphics{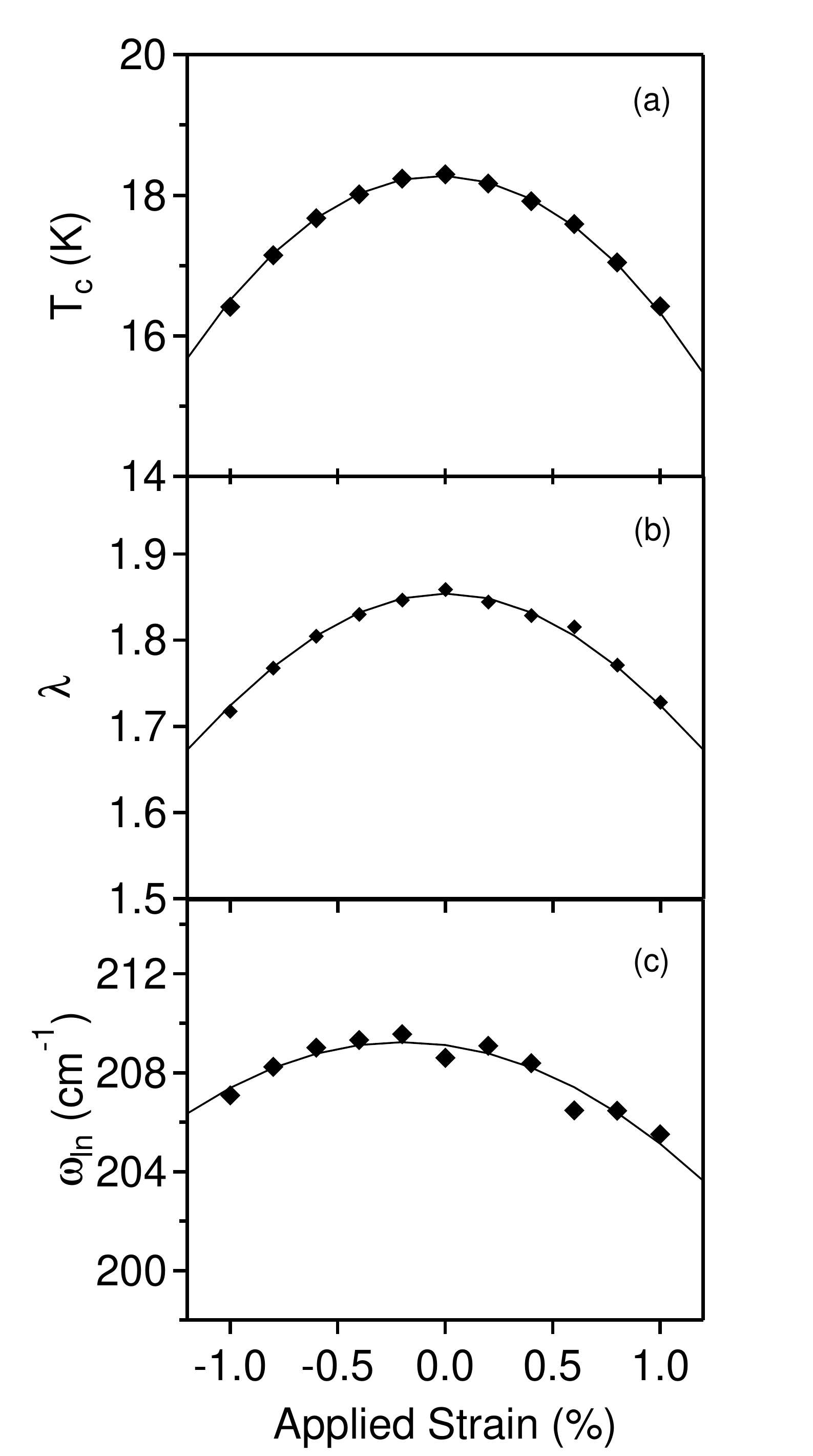}}
\caption{\label{fig:4}The behavior of: (a) the superconducting critical temperature, $T_c$; (b) the \emph{el-ph} coupling, $\lambda$; and (c) the logarithmically averaged phonon frequency $\omega{_ln}$ as a function of an applied uniaxial strain. Lines are guide for eye.}
\end{center}
\end{figure}

The calculated electronic structure along many high-symmetry directions of the simple-cubic Brillouin zone are displayed in Fig. \ref{fig:2} for three representative values of the applied strain (zero, 1.0\% and -1.0\%). The energy bands of the cubic crystal are shown as diamonds, whereas the circles and the triangles represents the 1.0\% and -1.0\% strain states, respectively. Indeed, the tetragonal deformation does not affect the electronic structure in a severe way: the energy bands of the distorted and undistorted cell are almost unchanged, and there is no evident splitting of the cubic bands at the Fermi level, $E_F$. The electronic DOS is also not drastically affected by the tetragonal distortion. In Fig. \ref{fig:2}, the Fermi level is marked by a dashed horizontal line and is set to 0 eV. It is interesting to notice that $E_F$ falls close to a sharp peak in the electronic DOS \cite{sadigh98}, with a value for the density of states of the order of 20 states$\slash$eV. This peak is generated by several nearly dispersionless bands crossing the Fermi level in the $\Gamma-M$, $\Gamma-R$, and $M-R$  directions and deriving from the \emph{4d} states of Nb atoms \cite{tutuncu06}.

\begin{figure}[t]
\begin{center}
\scalebox{0.47}{\includegraphics{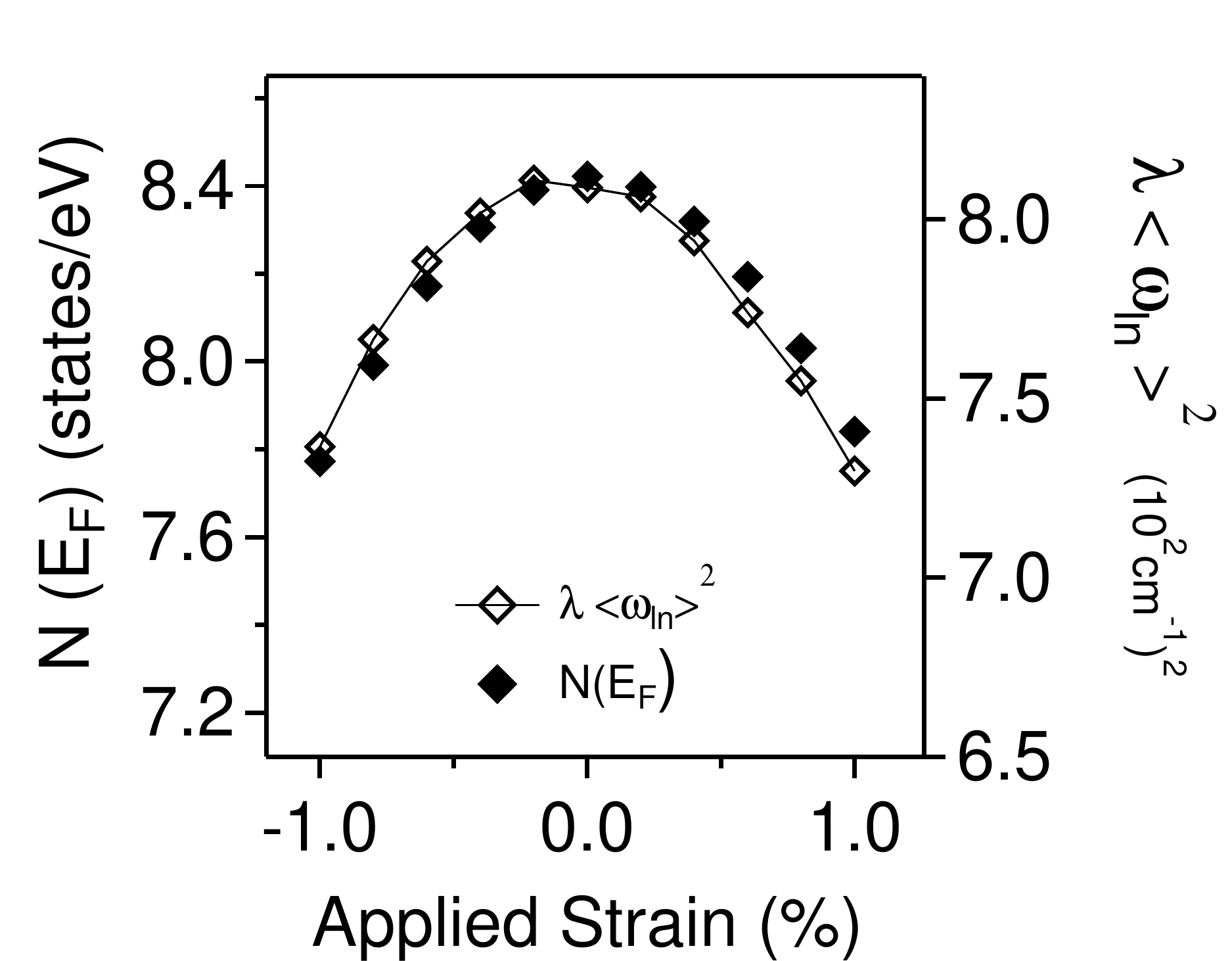}}
\caption{\label{fig:5}Density of states at Fermi level, $N(E_F)$ calculated as a function of strain. The curve is overlapped to the product function $\lambda\omega_{ln}^2$, which in good approximation should be proportional to $N(E_F)$\cite{mcmillan68}.}
\end{center}
\end{figure}

The phonon dispersion curves, calculated  along several high-symmetry directions in the Brillouin zone, are plotted for three representative strains in Fig. \ref{fig:3}; the phonon DOS is depicted in the right panel. There is good agreement with previous published results \cite{tutuncu06} at \emph{$\epsilon$ = 0}. For the sake of completeness, some inelastic neutron scattering measurements \cite{pintschovius85, axe65, shirane78} are also reported in Fig. \ref{fig:3}a, and again a good agreement is obtained.

\subsection{Derivation of superconducting $T_c$ as a function of strain}

The modifications of the phonon dispersion curves induced by a tetragonal distortion should have a strong effect on \emph{$T_c$} through the strain sensitivities of the averaged phonon frequencies $\omega_{ln}$, and of the \emph{el-ph} coupling $\lambda$. Therefore, both $\omega_{ln}(\epsilon)$ and $\lambda(\epsilon)$ have been explicitly calculated, and $T_c(\epsilon)$ has been estimated by means of the Allen-Dynes modification of the McMillan formula \cite{mcmillan68, allen75}:

\begin{equation}
   T_c = \frac{\hbar\omega_{ln}}{1.20}e^{-\frac{1.04(1+\lambda)}{\lambda - \mu^*(1 + 0.62\lambda)}}
\end{equation}

where $\mu^*$ is the effective Coulomb-repulsion parameter which describes the interaction beetween electrons, and $\omega_{ln}$ is a weighted logaritmically averaged phonon frequency, defined as:

\begin{equation}
   \omega_{ln} = e^{\frac{2}{\lambda}\int_0^{+\infty} \frac{\mathrm{d}\omega}{\omega} \! \alpha^2(\omega)F(\omega)\mathrm{ln}\omega \,}
\end{equation}

where $\alpha^2(\omega)F(\omega)$ is the Eliashberg spectral function. We assumed a negligible strain dependence of $\mu^*$ compared to the other parameters, and frozen it as a constant in our computations \cite{taylor05}.
Our results are reported in Fig. \ref{fig:4}a-c. 
As far as the cubic phase is concerned, our finding are in good agreement with the ones by T\"ut\"unc\"u \emph{et al.} \cite{tutuncu06}: a group of six phonon modes at the R-point (whose averaged phonon frequency is approximately 140 cm$^{-1}$) are found to strongly interact with the \emph{p-d} electronic states near the Fermi level. In these modes only the Nb chains vibrate, the Sn atoms being frozen at their equilibrium positions. The $\lambda_{\textbf{q}j}$ corresponding to these modes are comprised in the range 0.134$-$0.197. The overall electron-phonon interaction parameter ($\lambda$ = 1.85) agrees with the experimentally measured value \cite{wolf80}, and - choosing $\mu^*$ = 0.25 - the estimated critical temperature for the strain-free state is $T_c$ = 18.3 K (also in agreement with the highest reported $T_c$ \cite{hanak64}).

As it can be clearly seen in Figs. \ref{fig:4}b and \ref{fig:4}c, both  $\omega_{ln}(\epsilon)$ and $\lambda(\epsilon)$ show a parabolic profile as a function of strain. Very close to the cubic phase (strain-free cell), $\omega_{ln}$ has a maximum, implying a softening of the logarithmically averaged phonon frequencies when the system undergoes a distortion. The same behavior is found for $\lambda$, whose maximum is $\sim$ 1.85. The strength of the \emph{el-ph} interaction weakens as $\vert\epsilon\vert$ increases. As a result, by varying the axial strain the $T_c$ curve assumes the characteristic bell shape (Fig. \ref{fig:4}a) \cite{flukiger84,ekin07}. In addition, the curve shows a slight asymmetry with respect to the maximum, due mainly to an asymmetry in the phononic contribution ($\omega_{ln}$). However, a clear confirmation of this would deserve a more detailed analysis, based on an increased density of points on the curve.
Qualitatively, the $T_c$ vs. $\epsilon$ curve reproduces the experimental strain sensitivity of the critical current found in A15 superconductors: as it is well known, when a Nb$_3$Sn multifilamentary wire is subjected to a longitudinal strain, its critical current shows a maximum at zero intrinsic strain, and decreases reversibly with the applied load (see for example Ref. \onlinecite{godeke_thesis} and references therein), with a slight asymmetry\cite{flukiger05} between the compressive and tensile sides. In this sense, these first-principle calculations suggest that the origin of such strain sensitivity in Nb$_3$Sn is intrinsic and microscopic in its nature.

Our calculations also show that $N(E_F)$ is influenced by strain. This quantity is related to the \emph{el-ph} coupling constant and to the averaged phonon frequency trhough the following expression \cite{mcmillan68}: 

\begin{equation}\label{eq:dos}
   N(E_F) = M <I^2>\omega_{RMS}^2\lambda
\end{equation}

\begin{figure}[ht]
\begin{center}
\scalebox{0.3}{\includegraphics{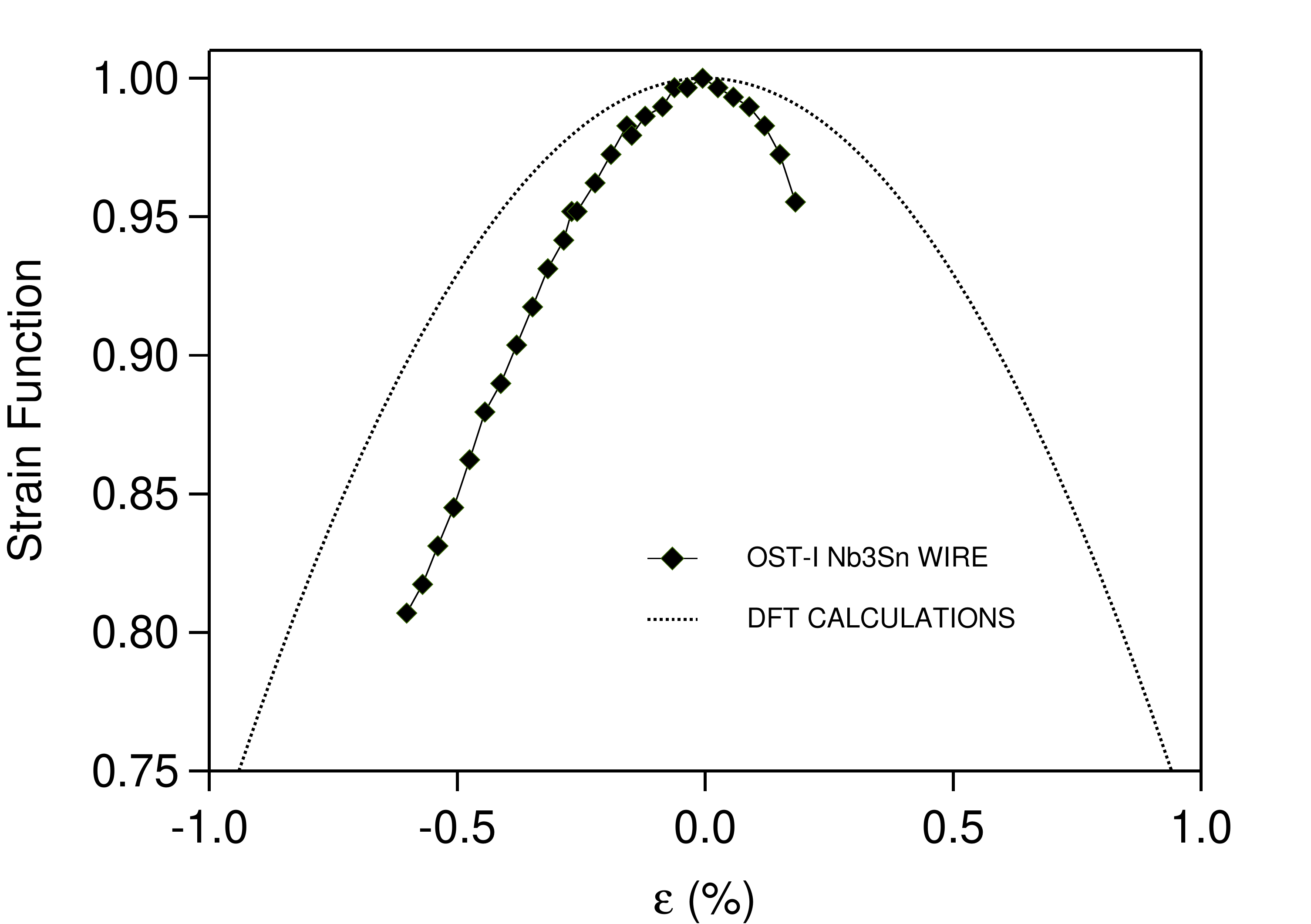}}
\caption{\label{fig:6}Direct comparison between the theoretical (dashed line) and experimental (line and markers) strain function\cite{demarzi12}. The theoretical $s(\epsilon)$ correctly reproduces a bell shape, the mismatch being attributed to those extrinsic effects generally observed in technological wires.}
\end{center}
\end{figure}

in which $<I^2>$ is the average over the Fermi surface of the \emph{el-ph} matrix element squared, $\omega_{RMS}$ is a weighted RMS phonon frequency and \emph{M} is the average ionic mass. By further assuming that the strain sensitivity of the normalized averaged frequencies $\omega_{RMS}$ and $\omega_{ln}$ are the same \cite{lim83} and that $<I^2>$ does not depend on any applied strain, it follows that $N(E_F, \epsilon)\propto\lambda\omega_{ln}^2$. The strain dependence of $N(E_F)$, calculated either directly or through Eq. \ref{eq:dos} are consistent with one another, as can be clearly seen in Fig. \ref{fig:5}.
Results of Figs.\ref{fig:4} and \ref{fig:5} thus provide evidence that the strain is affecting both the phononic and the electronic properties in the same way.
Indeed, existing studies consider only the two extreme cases where either the lattice deformations or the electronic properties modifications are considered as a source for the strain sensitivity. Markiewicz \cite{markiewicz04, markiewicz04bis, markiewicz06} has attempted to correlate the microscopic full invariant analysis with the Elisahberg-based relations for $T_c$ through strain-induced modifications in the electron-phonon spectrum. In this model, however, the strain induced changes in $N(E_F)$ are only accounted for through a strain-modified frequency dependence of the \emph{el-ph} interaction. In other words, the changes in $N(E_F)$ are not directly calculated, although the model is sufficiently accurate. Furthermore, microscopical theoretical predictions by Taylor and Hampshire \cite{taylor05} and Markiewicz \cite{markiewicz04bis} have shown that the variation of the superconducting properties of Nb$_3$Sn multifilamentary wires submitted to uniaxial strain are correlated to changes of the phonon spectrum,  rather than to the electronic density of states. On the other side, many works\cite{weger74, klein79bis, lim83} have linked the superconducting properties of A15 compounds to the variations in the electronic properties and $N(E_F)$.

 From the experimental point of view, it is important to highlight new methods for extracting the electron DOS from resistivity data have been explored with the aim to study whether the strain sensitivity is correlated to the electronic modifications\cite{mentink12}.

However, according to our results, any model aiming to describe the superconducting properties of A15 compounds from microscopic theories should take both contributions into account.

\subsection{Comparison with experimental results}

Starting from the pioneering work of Ekin \cite{ekin80}, in many models available in literature \cite{ekin80, taylor05, oh06, godeke06, arbelaez09, markiewicz06} the strain sensitivity of Nb$_3$Sn is generally parameterized using the strain function, s($\epsilon$), defined as \cite{ekin80, welch80}:

\begin{equation}\label{eq:strainfunction}
   s(\epsilon) \doteq \frac{B_{c2}(\epsilon, T = 0)}{B_{c2}(0, T = 0)} =  \left[ \frac{T_c(\epsilon)}{T_c(0)}\right]^w
\end{equation}

where $B_{c2}(\epsilon, T)$ represents the superconducting upper critical field, depending on strain and temperature, and where \emph{w} $\approx$ 3  for A15 materials \cite{ekin80}.

The strain function calculated using Eq. \ref{eq:strainfunction} is plotted in Fig. \ref{fig:6} (dashed line) and compared with the curve extracted from experimental $I_c vs. \epsilon$ measurements \cite{demarzi12} on a technological, multifilamentary wire (lines and markers in the figure). Although both curves show the expected bell shape, the experimental and theoretical curves are quite different. In particular, the strain sensitivity of the composite wire is enhanced when compared to the theoretical prediction of a bulk, perfectly binary and stoichiometric Nb$_3$Sn system, and this remains valid also if the value of \emph{w} is increased (within physical accepted ranges). This is not surprising, and the difference might have either an intrinsic or extrinsic origin. Among extrinsic phenomena inducing performance degradation with strain in technological wires\cite{ekin77}, filament breakage, reduction in wire's cross-sectional area, stress-induced martensitic formation\cite{demarzi12}, and microcrack/defect formation in the superconductor might play a role. The reversibility of the experimental data plotted in Fig. \ref{fig:6} implies that some of these phenomena can be ruled out, \emph{e.g.} filament breakage, microcrack and extended defect formation. All others mentioned above, being reversible, can in principle explain the strain sensitivity in Nb$_3$Sn, but their effect over $I_c$ is small and cannot account for the observed behavior \cite{ekin77}.
As far as intrinsic mechanisms are considered, our calculations unambiguously show that strain sensitivity in Nb$_3$Sn is associated to lattice and electronic deformations, which result in shifts in the Nb$_3$Sn critical surface.

It should be underlined the fact that in this first principles study we have neglected any sublattice displacements of the Nb atoms \cite{sadigh98} leading to Nb-chains dimerization. This can also have an effect on the theoretical $s(\epsilon)$. In our calculations, Nb atoms are frozen in their ideal positions, and the cubic cell is stable with respect to a tetragonal strain\cite{sadigh98}. However, if the Nb atoms are allowed to relax, a dimerization of the chains occurs, and the undistorted cubic structure become unstable with respect to a spontaneous sublattice distortion.  However, such distortions are small (the ratio between the major and minor axis of the tetragonal cell spans from 0.9938 to 0.9964\cite{sadigh98}) and therefore their effect on $s(\epsilon)$ is expected to be negligible.
Also, our system is perfectly binary, whereas in technological wires Ti or Ta is inserted as the ternary element, with the aim to improve the pinning efficiency and therefore $J_c(B)$. Moreover, due to compositional inhomogeneities, a distribution of the superconducting properties is observed ($T_c$ depends on the atomic Sn content, having a maximum at Sn\% = 0.25\%); see for example Refs. \onlinecite{senatore07} and \onlinecite{godeke06bis} and reference therein.
Considering all these aspects, it is clear that the polycrystalline Nb$_3$Sn formed by a reaction heat treatment inside a composite, multifilamentary system has a microscopic structure that inevitably deviates from that of an ideal Nb$_3$Sn lattice cell. Therefore, differences between the theoretical computations of $s(\epsilon)$ and the experimental degradation in wires can be expected, which might possibly suggest paths towards the improvement of the strain tolerance in technological wires. For example, at 0.5\% compressive strain the strain sensitivity could in principle be reduced to $\sim$ 0.07\%, thus helping in the design of those devices where the levels of strain experienced by Nb$_3$Sn are sufficiently large, as is the case of high-field superconducting magnets.

%\section{CONCLUSION}

\begin{acknowledgments}
We thank Federico Quagliata and Pietro D'Angelo for setting up the environment and the CRESCO parallel cluster at ENEA C.R. Frascati. M.B.N. wishes to acknowledge partial support from the Office of Basic Energy Sciences, U.S. Department of Energy at Oak Ridge National Laboratory under Contract No. DE-AC05-00OR22725 with UT-Battelle, LLC.
\end{acknowledgments}

% Create the reference section using BibTeX:
\bibliography{nb3sn_strain}

\begin{thebibliography}{62}
\expandafter\ifx\csname natexlab\endcsname\relax\def\natexlab#1{#1}\fi
\expandafter\ifx\csname bibnamefont\endcsname\relax
  \def\bibnamefont#1{#1}\fi
\expandafter\ifx\csname bibfnamefont\endcsname\relax
  \def\bibfnamefont#1{#1}\fi
\expandafter\ifx\csname citenamefont\endcsname\relax
  \def\citenamefont#1{#1}\fi
\expandafter\ifx\csname url\endcsname\relax
  \def\url#1{\texttt{#1}}\fi
\expandafter\ifx\csname urlprefix\endcsname\relax\def\urlprefix{URL }\fi
\providecommand{\bibinfo}[2]{#2}
\providecommand{\eprint}[2][]{\url{#2}}

\bibitem[{\citenamefont{Matthias}(1953)}]{Matthias53}
\bibinfo{author}{\bibfnamefont{B.~T.} \bibnamefont{Matthias}},
  \bibinfo{journal}{Phys. Rev.} \textbf{\bibinfo{volume}{92}},
  \bibinfo{pages}{874} (\bibinfo{year}{1953}).

\bibitem[{\citenamefont{Miyazaki et~al.}(2003)\citenamefont{Miyazaki, Hase, and
  Miyatake}}]{miyazaki03}
\bibinfo{author}{\bibfnamefont{T.}~\bibnamefont{Miyazaki}},
  \bibinfo{author}{\bibfnamefont{T.}~\bibnamefont{Hase}}, \bibnamefont{and}
  \bibinfo{author}{\bibfnamefont{T.}~\bibnamefont{Miyatake}},
  \emph{\bibinfo{title}{Handbook of Superconducting Materials}}
  (\bibinfo{publisher}{Bristol: IOP Publishing}, \bibinfo{year}{2003}), pp.
  \bibinfo{pages}{639--72}.

\bibitem[{\citenamefont{Vostner and Salpietro}(2006)}]{vostner06}
\bibinfo{author}{\bibfnamefont{A.}~\bibnamefont{Vostner}} \bibnamefont{and}
  \bibinfo{author}{\bibfnamefont{E.}~\bibnamefont{Salpietro}},
  \bibinfo{journal}{Supercond. Sci. Technol.} \textbf{\bibinfo{volume}{19}},
  \bibinfo{pages}{S90} (\bibinfo{year}{2006}).

\bibitem[{\citenamefont{Mitchell et~al.}(2012)\citenamefont{Mitchell, Devred,
  Libeyre, Lim, F., and the ITER Magnet~Division}}]{mitchell12}
\bibinfo{author}{\bibfnamefont{N.}~\bibnamefont{Mitchell}},
  \bibinfo{author}{\bibfnamefont{A.}~\bibnamefont{Devred}},
  \bibinfo{author}{\bibfnamefont{P.}~\bibnamefont{Libeyre}},
  \bibinfo{author}{\bibfnamefont{B.}~\bibnamefont{Lim}},
  \bibinfo{author}{\bibfnamefont{S.}~\bibnamefont{F.}}, \bibnamefont{and}
  \bibinfo{author}{\bibnamefont{the ITER Magnet~Division}},
  \bibinfo{journal}{IEEE Trans. on Appl. Supercond.}
  \textbf{\bibinfo{volume}{22}}, \bibinfo{pages}{4200809}
  (\bibinfo{year}{2012}).

\bibitem[{\citenamefont{Devred et~al.}(2012)\citenamefont{Devred, Backbier,
  Bessette, Bevillard, Gardner, Jewell, Mitchell, Pong, and
  Vostner}}]{devred12}
\bibinfo{author}{\bibfnamefont{A.}~\bibnamefont{Devred}},
  \bibinfo{author}{\bibfnamefont{I.}~\bibnamefont{Backbier}},
  \bibinfo{author}{\bibfnamefont{D.}~\bibnamefont{Bessette}},
  \bibinfo{author}{\bibfnamefont{G.}~\bibnamefont{Bevillard}},
  \bibinfo{author}{\bibfnamefont{M.}~\bibnamefont{Gardner}},
  \bibinfo{author}{\bibfnamefont{M.}~\bibnamefont{Jewell}},
  \bibinfo{author}{\bibfnamefont{N.}~\bibnamefont{Mitchell}},
  \bibinfo{author}{\bibfnamefont{I.}~\bibnamefont{Pong}}, \bibnamefont{and}
  \bibinfo{author}{\bibfnamefont{A.}~\bibnamefont{Vostner}},
  \bibinfo{journal}{IEEE Trans. on Appl. Supercond.}
  \textbf{\bibinfo{volume}{22}}, \bibinfo{pages}{4804909}
  (\bibinfo{year}{2012}).

\bibitem[{\citenamefont{Wada and Kiyoshi}(2002)}]{wada02}
\bibinfo{author}{\bibfnamefont{H.}~\bibnamefont{Wada}} \bibnamefont{and}
  \bibinfo{author}{\bibfnamefont{T.}~\bibnamefont{Kiyoshi}},
  \bibinfo{journal}{IEEE Trans. Appl. Supercond.}
  \textbf{\bibinfo{volume}{12}}, \bibinfo{pages}{715} (\bibinfo{year}{2002}).

\bibitem[{\citenamefont{Bottura et~al.}(2012)\citenamefont{Bottura, de~Rijk,
  Rossi, and Todesco}}]{bottura12}
\bibinfo{author}{\bibfnamefont{L.}~\bibnamefont{Bottura}},
  \bibinfo{author}{\bibfnamefont{G.}~\bibnamefont{de~Rijk}},
  \bibinfo{author}{\bibfnamefont{L.}~\bibnamefont{Rossi}}, \bibnamefont{and}
  \bibinfo{author}{\bibfnamefont{E.}~\bibnamefont{Todesco}},
  \bibinfo{journal}{IEEE Trans. on Appl. Supercond.}
  \textbf{\bibinfo{volume}{22}}, \bibinfo{pages}{4002008}
  (\bibinfo{year}{2012}).

\bibitem[{\citenamefont{Rupp}(1977)}]{rupp77}
\bibinfo{author}{\bibfnamefont{G.}~\bibnamefont{Rupp}}, \bibinfo{journal}{J.
  Appl. Phys.} \textbf{\bibinfo{volume}{48}}, \bibinfo{pages}{3858}
  (\bibinfo{year}{1977}).

\bibitem[{\citenamefont{Ekin}(1979)}]{ekin79}
\bibinfo{author}{\bibfnamefont{J.}~\bibnamefont{Ekin}}, \bibinfo{journal}{IEEE
  Trans. Magn.} \textbf{\bibinfo{volume}{15}}, \bibinfo{pages}{197}
  (\bibinfo{year}{1979}).

\bibitem[{\citenamefont{ten Haken}(1994)}]{tenhaken94}
\bibinfo{author}{\bibfnamefont{B.}~\bibnamefont{ten Haken}},
  \bibinfo{type}{{Ph.D. Thesis}}, \bibinfo{school}{{Technical University of
  Twente, Enschede, The Netherlands}} (\bibinfo{year}{1994}).

\bibitem[{\citenamefont{Ekin}(1987)}]{ekin87}
\bibinfo{author}{\bibfnamefont{J.}~\bibnamefont{Ekin}}, \bibinfo{journal}{J.
  Appl. Phys.} \textbf{\bibinfo{volume}{62}}, \bibinfo{pages}{4829}
  (\bibinfo{year}{1987}).

\bibitem[{\citenamefont{Hoenig et~al.}(1979)\citenamefont{Hoenig, Montgomery,
  and Waldman}}]{hoenig79}
\bibinfo{author}{\bibfnamefont{M.~O.} \bibnamefont{Hoenig}},
  \bibinfo{author}{\bibfnamefont{A.~G.} \bibnamefont{Montgomery}},
  \bibnamefont{and} \bibinfo{author}{\bibfnamefont{S.~J.}
  \bibnamefont{Waldman}}, \bibinfo{journal}{IEEE Trans. on Mag.}
  \textbf{\bibinfo{volume}{15}}, \bibinfo{pages}{792} (\bibinfo{year}{1979}).

\bibitem[{\citenamefont{Bruzzone}(2006)}]{bruzzone06}
\bibinfo{author}{\bibfnamefont{P.}~\bibnamefont{Bruzzone}},
  \bibinfo{journal}{IEEE Trans. Appl. Supercond.}
  \textbf{\bibinfo{volume}{16}}, \bibinfo{pages}{839} (\bibinfo{year}{2006}).

\bibitem[{\citenamefont{Spadoni}(1994)}]{spadoni94}
\bibinfo{author}{\bibfnamefont{M.}~\bibnamefont{Spadoni}},
  \bibinfo{journal}{IEEE Trans. on Magn.} \textbf{\bibinfo{volume}{30}},
  \bibinfo{pages}{1699} (\bibinfo{year}{1994}).

\bibitem[{\citenamefont{Nijhuis and Ilyin}(2006)}]{nijhuis06}
\bibinfo{author}{\bibfnamefont{A.}~\bibnamefont{Nijhuis}} \bibnamefont{and}
  \bibinfo{author}{\bibfnamefont{Y.}~\bibnamefont{Ilyin}},
  \bibinfo{journal}{Supercond. Sci. Technol.} \textbf{\bibinfo{volume}{19}},
  \bibinfo{pages}{945} (\bibinfo{year}{2006}).

\bibitem[{\citenamefont{Mitchell}(2005)}]{mitchell05}
\bibinfo{author}{\bibfnamefont{N.}~\bibnamefont{Mitchell}},
  \bibinfo{journal}{Supercond. Sci. Technol.} \textbf{\bibinfo{volume}{18}},
  \bibinfo{pages}{S396} (\bibinfo{year}{2005}).

\bibitem[{\citenamefont{Mondonico et~al.}(2012)\citenamefont{Mondonico, Seeber,
  Ferreira, Bordini, Oberli, Bottura, Ballarino, Fl\"ukiger, and
  Senatore}}]{mondonico12}
\bibinfo{author}{\bibfnamefont{G.}~\bibnamefont{Mondonico}},
  \bibinfo{author}{\bibfnamefont{B.}~\bibnamefont{Seeber}},
  \bibinfo{author}{\bibfnamefont{A.}~\bibnamefont{Ferreira}},
  \bibinfo{author}{\bibfnamefont{B.}~\bibnamefont{Bordini}},
  \bibinfo{author}{\bibfnamefont{L.}~\bibnamefont{Oberli}},
  \bibinfo{author}{\bibfnamefont{L.}~\bibnamefont{Bottura}},
  \bibinfo{author}{\bibfnamefont{A.}~\bibnamefont{Ballarino}},
  \bibinfo{author}{\bibfnamefont{R.}~\bibnamefont{Fl\"ukiger}},
  \bibnamefont{and} \bibinfo{author}{\bibfnamefont{C.}~\bibnamefont{Senatore}},
  \bibinfo{journal}{Supercond. Sci. Technol.} \textbf{\bibinfo{volume}{25}},
  \bibinfo{pages}{115002} (\bibinfo{year}{2012}).

\bibitem[{\citenamefont{Ekin}(1980)}]{ekin80}
\bibinfo{author}{\bibfnamefont{J.~W.} \bibnamefont{Ekin}},
  \bibinfo{journal}{Cryogenics} \textbf{\bibinfo{volume}{20}},
  \bibinfo{pages}{611} (\bibinfo{year}{1980}).

\bibitem[{\citenamefont{Taylor and Hampshire}(2005)}]{taylor05}
\bibinfo{author}{\bibfnamefont{D.~M.} \bibnamefont{Taylor}} \bibnamefont{and}
  \bibinfo{author}{\bibfnamefont{D.~P.} \bibnamefont{Hampshire}},
  \bibinfo{journal}{Supercond. Sci. Technol.} \textbf{\bibinfo{volume}{18}},
  \bibinfo{pages}{S241} (\bibinfo{year}{2005}).

\bibitem[{\citenamefont{Oh and Kim}(2006)}]{oh06}
\bibinfo{author}{\bibfnamefont{S.}~\bibnamefont{Oh}} \bibnamefont{and}
  \bibinfo{author}{\bibfnamefont{K.}~\bibnamefont{Kim}}, \bibinfo{journal}{J.
  Appl. Phys.} \textbf{\bibinfo{volume}{99}}, \bibinfo{pages}{0330909}
  (\bibinfo{year}{2006}).

\bibitem[{\citenamefont{Godeke et~al.}(2006)\citenamefont{Godeke, ten Katen,
  ten Kate, and Larbalestier}}]{godeke06}
\bibinfo{author}{\bibfnamefont{A.}~\bibnamefont{Godeke}},
  \bibinfo{author}{\bibfnamefont{B.}~\bibnamefont{ten Katen}},
  \bibinfo{author}{\bibfnamefont{H.~H.~J.} \bibnamefont{ten Kate}},
  \bibnamefont{and} \bibinfo{author}{\bibfnamefont{D.~C.}
  \bibnamefont{Larbalestier}}, \bibinfo{journal}{Supercond. Sci. Technol.}
  \textbf{\bibinfo{volume}{19}}, \bibinfo{pages}{R100} (\bibinfo{year}{2006}).

\bibitem[{\citenamefont{Arbelaez et~al.}(2009)\citenamefont{Arbelaez, Godeke,
  and Prestemon}}]{arbelaez09}
\bibinfo{author}{\bibfnamefont{D.}~\bibnamefont{Arbelaez}},
  \bibinfo{author}{\bibfnamefont{A.}~\bibnamefont{Godeke}}, \bibnamefont{and}
  \bibinfo{author}{\bibfnamefont{S.}~\bibnamefont{Prestemon}},
  \bibinfo{journal}{Supercond. Sci. Technol.} \textbf{\bibinfo{volume}{22}}
  (\bibinfo{year}{2009}).

\bibitem[{\citenamefont{Markiewicz}(2006)}]{markiewicz06}
\bibinfo{author}{\bibfnamefont{W.~D.} \bibnamefont{Markiewicz}},
  \bibinfo{journal}{Cryogenics} \textbf{\bibinfo{volume}{46}},
  \bibinfo{pages}{864} (\bibinfo{year}{2006}).

\bibitem[{\citenamefont{Markiewicz}(2004{\natexlab{a}})}]{markiewicz04}
\bibinfo{author}{\bibfnamefont{W.~D.} \bibnamefont{Markiewicz}},
  \bibinfo{journal}{Cryogenics} \textbf{\bibinfo{volume}{44}},
  \bibinfo{pages}{767} (\bibinfo{year}{2004}{\natexlab{a}}).

\bibitem[{\citenamefont{Muzzi et~al.}(2012)\citenamefont{Muzzi, Corato, della
  Corte, Marzi, Spina, Daniels, Michiel, Buta, Mondonico, Seeber
  et~al.}}]{muzzi12}
\bibinfo{author}{\bibfnamefont{L.}~\bibnamefont{Muzzi}},
  \bibinfo{author}{\bibfnamefont{V.}~\bibnamefont{Corato}},
  \bibinfo{author}{\bibfnamefont{A.}~\bibnamefont{della Corte}},
  \bibinfo{author}{\bibfnamefont{G.~D.} \bibnamefont{Marzi}},
  \bibinfo{author}{\bibfnamefont{T.}~\bibnamefont{Spina}},
  \bibinfo{author}{\bibfnamefont{J.}~\bibnamefont{Daniels}},
  \bibinfo{author}{\bibfnamefont{M.~D.} \bibnamefont{Michiel}},
  \bibinfo{author}{\bibfnamefont{F.}~\bibnamefont{Buta}},
  \bibinfo{author}{\bibfnamefont{G.}~\bibnamefont{Mondonico}},
  \bibinfo{author}{\bibfnamefont{B.}~\bibnamefont{Seeber}},
  \bibnamefont{et~al.}, \bibinfo{journal}{Supercond. Sci. Technol.}
  \textbf{\bibinfo{volume}{25}}, \bibinfo{pages}{054006}
  (\bibinfo{year}{2012}).

\bibitem[{\citenamefont{Klein and Lu}(2001)}]{klein01}
\bibinfo{author}{\bibfnamefont{B.~M.} \bibnamefont{Klein}} \bibnamefont{and}
  \bibinfo{author}{\bibfnamefont{Z.}~\bibnamefont{Lu}},
  \bibinfo{journal}{Physica B: Condensed Matter}
  \textbf{\bibinfo{volume}{296}}, \bibinfo{pages}{120 } (\bibinfo{year}{2001}).

\bibitem[{\citenamefont{Sadigh and Ozoli\c{n}\v{s}}(1998)}]{sadigh98}
\bibinfo{author}{\bibfnamefont{B.}~\bibnamefont{Sadigh}} \bibnamefont{and}
  \bibinfo{author}{\bibfnamefont{V.}~\bibnamefont{Ozoli\c{n}\v{s}}},
  \bibinfo{journal}{Phys. Rev. B} \textbf{\bibinfo{volume}{57}},
  \bibinfo{pages}{2793} (\bibinfo{year}{1998}).

\bibitem[{\citenamefont{Lu and Klein}(1997)}]{lu97}
\bibinfo{author}{\bibfnamefont{Z.~W.} \bibnamefont{Lu}} \bibnamefont{and}
  \bibinfo{author}{\bibfnamefont{B.~M.} \bibnamefont{Klein}},
  \bibinfo{journal}{Phys. Rev. Lett.} \textbf{\bibinfo{volume}{79}},
  \bibinfo{pages}{1361} (\bibinfo{year}{1997}).

\bibitem[{\citenamefont{Mattheiss and Weber}(1982)}]{mattheiss82}
\bibinfo{author}{\bibfnamefont{L.~F.} \bibnamefont{Mattheiss}}
  \bibnamefont{and} \bibinfo{author}{\bibfnamefont{W.}~\bibnamefont{Weber}},
  \bibinfo{journal}{Phys. Rev. B} \textbf{\bibinfo{volume}{25}},
  \bibinfo{pages}{2248} (\bibinfo{year}{1982}).

\bibitem[{\citenamefont{Klein et~al.}(1978{\natexlab{a}})\citenamefont{Klein,
  Boyer, Papaconstantopoulos, and Mattheiss}}]{klein78}
\bibinfo{author}{\bibfnamefont{B.~M.} \bibnamefont{Klein}},
  \bibinfo{author}{\bibfnamefont{L.~L.} \bibnamefont{Boyer}},
  \bibinfo{author}{\bibfnamefont{D.~A.} \bibnamefont{Papaconstantopoulos}},
  \bibnamefont{and} \bibinfo{author}{\bibfnamefont{L.~F.}
  \bibnamefont{Mattheiss}}, \bibinfo{journal}{Phys. Rev. B}
  \textbf{\bibinfo{volume}{18}}, \bibinfo{pages}{6411}
  (\bibinfo{year}{1978}{\natexlab{a}}).

\bibitem[{\citenamefont{Mattheiss}(1975)}]{mattheiss75}
\bibinfo{author}{\bibfnamefont{L.~F.} \bibnamefont{Mattheiss}},
  \bibinfo{journal}{Phys. Rev. B} \textbf{\bibinfo{volume}{12}},
  \bibinfo{pages}{2161} (\bibinfo{year}{1975}).

\bibitem[{\citenamefont{Weber}(1984{\natexlab{a}})}]{weber84}
\bibinfo{author}{\bibfnamefont{W.}~\bibnamefont{Weber}},
  \bibinfo{journal}{Physica B+C} \textbf{\bibinfo{volume}{126}},
  \bibinfo{pages}{217 } (\bibinfo{year}{1984}{\natexlab{a}}), ISSN
  \bibinfo{issn}{0378-4363}.

\bibitem[{\citenamefont{Weber}(1984{\natexlab{b}})}]{weber84book}
\bibinfo{author}{\bibfnamefont{W.}~\bibnamefont{Weber}},
  \emph{\bibinfo{title}{Electronic Structure of Complex Systems}}
  (\bibinfo{publisher}{Plenum Press, New York},
  \bibinfo{year}{1984}{\natexlab{b}}), vol. \bibinfo{volume}{113}, p.
  \bibinfo{pages}{345}.

\bibitem[{\citenamefont{T\"ut\"unc\"u et~al.}(2006)\citenamefont{T\"ut\"unc\"u,
  Srivastava, Ba\ifmmode \breve{g}\else \u{g}\fi{}c\ifmmode \imath \else~\i
  \fi{}, and Duman}}]{tutuncu06}
\bibinfo{author}{\bibfnamefont{H.~M.} \bibnamefont{T\"ut\"unc\"u}},
  \bibinfo{author}{\bibfnamefont{G.~P.} \bibnamefont{Srivastava}},
  \bibinfo{author}{\bibfnamefont{S.}~\bibnamefont{Ba\ifmmode \breve{g}\else
  \u{g}\fi{}c\ifmmode \imath \else~\i \fi{}}}, \bibnamefont{and}
  \bibinfo{author}{\bibfnamefont{S.}~\bibnamefont{Duman}},
  \bibinfo{journal}{Phys. Rev. B} \textbf{\bibinfo{volume}{74}},
  \bibinfo{pages}{212506} (\bibinfo{year}{2006}).

\bibitem[{\citenamefont{Baroni et~al.}(1987)\citenamefont{Baroni, Giannozzi,
  and Testa}}]{baroni87}
\bibinfo{author}{\bibfnamefont{S.}~\bibnamefont{Baroni}},
  \bibinfo{author}{\bibfnamefont{P.}~\bibnamefont{Giannozzi}},
  \bibnamefont{and} \bibinfo{author}{\bibfnamefont{A.}~\bibnamefont{Testa}},
  \bibinfo{journal}{Phys. Rev. Lett.} \textbf{\bibinfo{volume}{58}},
  \bibinfo{pages}{1861} (\bibinfo{year}{1987}).

\bibitem[{\citenamefont{Baroni et~al.}(2001)\citenamefont{Baroni, de~Gironcoli,
  Dal~Corso, and Giannozzi}}]{baroni01}
\bibinfo{author}{\bibfnamefont{S.}~\bibnamefont{Baroni}},
  \bibinfo{author}{\bibfnamefont{S.}~\bibnamefont{de~Gironcoli}},
  \bibinfo{author}{\bibfnamefont{A.}~\bibnamefont{Dal~Corso}},
  \bibnamefont{and}
  \bibinfo{author}{\bibfnamefont{P.}~\bibnamefont{Giannozzi}},
  \bibinfo{journal}{Rev. Mod. Phys.} \textbf{\bibinfo{volume}{73}},
  \bibinfo{pages}{515} (\bibinfo{year}{2001}).

\bibitem[{QE()}]{QE}
\bibinfo{note}{{\sc Quantum-ESPRESSO} is a community project for high-quality
  quantum-simulation software, based on density-functional theory, and
  coordinated by Paolo Giannozzi. See http://www.quantum-espresso.org and
  http://www.pwscf.org.}

\bibitem[{\citenamefont{Troullier and Martins}(1991)}]{troullier91}
\bibinfo{author}{\bibfnamefont{N.}~\bibnamefont{Troullier}} \bibnamefont{and}
  \bibinfo{author}{\bibfnamefont{J.~L.} \bibnamefont{Martins}},
  \bibinfo{journal}{Phys. Rev. B} \textbf{\bibinfo{volume}{43}},
  \bibinfo{pages}{1993} (\bibinfo{year}{1991}).

\bibitem[{pse()}]{pseudo}
\bibinfo{note}{We used the pseudopotentials Nb.pw91-nsp-van.UPF and
  Sn.pw91-n-van.UPF from the http://www.quantum-espresso.org distribution.}

\bibitem[{\citenamefont{Maier}(1969)}]{maier69}
\bibinfo{author}{\bibfnamefont{R.~G.} \bibnamefont{Maier}},
  \bibinfo{journal}{Z. Naturforsch. Teil A} \textbf{\bibinfo{volume}{24}},
  \bibinfo{pages}{1033} (\bibinfo{year}{1969}).

\bibitem[{\citenamefont{Guritanu et~al.}(2004)\citenamefont{Guritanu,
  Goldacker, Bouquet, Wang, Lortz, Goll, and Junod}}]{guritanu04}
\bibinfo{author}{\bibfnamefont{V.}~\bibnamefont{Guritanu}},
  \bibinfo{author}{\bibfnamefont{W.}~\bibnamefont{Goldacker}},
  \bibinfo{author}{\bibfnamefont{F.}~\bibnamefont{Bouquet}},
  \bibinfo{author}{\bibfnamefont{Y.}~\bibnamefont{Wang}},
  \bibinfo{author}{\bibfnamefont{R.}~\bibnamefont{Lortz}},
  \bibinfo{author}{\bibfnamefont{G.}~\bibnamefont{Goll}}, \bibnamefont{and}
  \bibinfo{author}{\bibfnamefont{A.}~\bibnamefont{Junod}},
  \bibinfo{journal}{Phys. Rev. B} \textbf{\bibinfo{volume}{70}},
  \bibinfo{pages}{184526} (\bibinfo{year}{2004}).

\bibitem[{\citenamefont{McMillan}(1968)}]{mcmillan68}
\bibinfo{author}{\bibfnamefont{W.~L.} \bibnamefont{McMillan}},
  \bibinfo{journal}{Phys. Rev.} \textbf{\bibinfo{volume}{167}},
  \bibinfo{pages}{331} (\bibinfo{year}{1968}).

\bibitem[{\citenamefont{Pintschovius et~al.}(1985)\citenamefont{Pintschovius,
  Takei, and Toyota}}]{pintschovius85}
\bibinfo{author}{\bibfnamefont{L.}~\bibnamefont{Pintschovius}},
  \bibinfo{author}{\bibfnamefont{H.}~\bibnamefont{Takei}}, \bibnamefont{and}
  \bibinfo{author}{\bibfnamefont{N.}~\bibnamefont{Toyota}},
  \bibinfo{journal}{Phys. Rev. Lett.} \textbf{\bibinfo{volume}{54}},
  \bibinfo{pages}{1260} (\bibinfo{year}{1985}).

\bibitem[{\citenamefont{Axe and Shirane}(1973)}]{axe65}
\bibinfo{author}{\bibfnamefont{J.~D.} \bibnamefont{Axe}} \bibnamefont{and}
  \bibinfo{author}{\bibfnamefont{G.}~\bibnamefont{Shirane}},
  \bibinfo{journal}{Phys. Rev. B} \textbf{\bibinfo{volume}{8}},
  \bibinfo{pages}{1965} (\bibinfo{year}{1973}).

\bibitem[{\citenamefont{Shirane and Axe}(1978)}]{shirane78}
\bibinfo{author}{\bibfnamefont{G.}~\bibnamefont{Shirane}} \bibnamefont{and}
  \bibinfo{author}{\bibfnamefont{J.~D.} \bibnamefont{Axe}},
  \bibinfo{journal}{Phys. Rev. B} \textbf{\bibinfo{volume}{18}},
  \bibinfo{pages}{3742} (\bibinfo{year}{1978}).

\bibitem[{\citenamefont{Allen and Dynes}(1975)}]{allen75}
\bibinfo{author}{\bibfnamefont{P.~B.} \bibnamefont{Allen}} \bibnamefont{and}
  \bibinfo{author}{\bibfnamefont{R.~C.} \bibnamefont{Dynes}},
  \bibinfo{journal}{Phys. Rev. B} \textbf{\bibinfo{volume}{12}},
  \bibinfo{pages}{905} (\bibinfo{year}{1975}).

\bibitem[{\citenamefont{Wolf et~al.}(1980)\citenamefont{Wolf, Zasadzinski,
  Arnold, Moore, Rowell, and Beasley}}]{wolf80}
\bibinfo{author}{\bibfnamefont{E.~L.} \bibnamefont{Wolf}},
  \bibinfo{author}{\bibfnamefont{J.}~\bibnamefont{Zasadzinski}},
  \bibinfo{author}{\bibfnamefont{G.~B.} \bibnamefont{Arnold}},
  \bibinfo{author}{\bibfnamefont{D.~F.} \bibnamefont{Moore}},
  \bibinfo{author}{\bibfnamefont{J.~M.} \bibnamefont{Rowell}},
  \bibnamefont{and} \bibinfo{author}{\bibfnamefont{M.~R.}
  \bibnamefont{Beasley}}, \bibinfo{journal}{Phys. Rev. B}
  \textbf{\bibinfo{volume}{22}}, \bibinfo{pages}{1214} (\bibinfo{year}{1980}).

\bibitem[{\citenamefont{Hanak et~al.}(1964)\citenamefont{Hanak, Strater, and
  Cullen}}]{hanak64}
\bibinfo{author}{\bibfnamefont{J.}~\bibnamefont{Hanak}},
  \bibinfo{author}{\bibfnamefont{K.}~\bibnamefont{Strater}}, \bibnamefont{and}
  \bibinfo{author}{\bibfnamefont{R.}~\bibnamefont{Cullen}},
  \bibinfo{journal}{RCA Review} \textbf{\bibinfo{volume}{25}},
  \bibinfo{pages}{342} (\bibinfo{year}{1964}).

\bibitem[{\citenamefont{Fl\"ukiger et~al.}(1984)\citenamefont{Fl\"ukiger,
  Isernhage, Goldhacker, and Specking}}]{flukiger84}
\bibinfo{author}{\bibfnamefont{R.}~\bibnamefont{Fl\"ukiger}},
  \bibinfo{author}{\bibfnamefont{R.}~\bibnamefont{Isernhage}},
  \bibinfo{author}{\bibfnamefont{W.}~\bibnamefont{Goldhacker}},
  \bibnamefont{and} \bibinfo{author}{\bibfnamefont{W.}~\bibnamefont{Specking}},
  \bibinfo{journal}{Adv. Cryo. Eng. (Materials)} \textbf{\bibinfo{volume}{30}},
  \bibinfo{pages}{851} (\bibinfo{year}{1984}).

\bibitem[{\citenamefont{Ekin}(2007)}]{ekin07}
\bibinfo{author}{\bibfnamefont{J.~W.} \bibnamefont{Ekin}},
  \emph{\bibinfo{title}{Experimental Techniques for Low-Temperature
  Measurements}} (\bibinfo{publisher}{New York: Oxford University Press},
  \bibinfo{year}{2007}).

\bibitem[{\citenamefont{Godeke}(2005)}]{godeke_thesis}
\bibinfo{author}{\bibfnamefont{A.}~\bibnamefont{Godeke}}, \bibinfo{type}{{Ph.D.
  Thesis}}, \bibinfo{school}{{Technical University of Twente, Enschede, The
  Netherlands}} (\bibinfo{year}{2005}).

\bibitem[{\citenamefont{Fl\"ukiger et~al.}(2005)\citenamefont{Fl\"ukiger,
  Uglietti, Ab\"acherli, and Seeber}}]{flukiger05}
\bibinfo{author}{\bibfnamefont{R.}~\bibnamefont{Fl\"ukiger}},
  \bibinfo{author}{\bibfnamefont{D.}~\bibnamefont{Uglietti}},
  \bibinfo{author}{\bibfnamefont{V.}~\bibnamefont{Ab\"acherli}},
  \bibnamefont{and} \bibinfo{author}{\bibfnamefont{B.}~\bibnamefont{Seeber}},
  \bibinfo{journal}{Supercond. Sci. Technol.} \textbf{\bibinfo{volume}{18}},
  \bibinfo{pages}{S416} (\bibinfo{year}{2005}).

\bibitem[{\citenamefont{Marzi et~al.}(2012)\citenamefont{Marzi, Corato, Muzzi,
  della Corte, Mondonico, Seeber, and Senatore}}]{demarzi12}
\bibinfo{author}{\bibfnamefont{G.~D.} \bibnamefont{Marzi}},
  \bibinfo{author}{\bibfnamefont{V.}~\bibnamefont{Corato}},
  \bibinfo{author}{\bibfnamefont{L.}~\bibnamefont{Muzzi}},
  \bibinfo{author}{\bibfnamefont{A.}~\bibnamefont{della Corte}},
  \bibinfo{author}{\bibfnamefont{G.}~\bibnamefont{Mondonico}},
  \bibinfo{author}{\bibfnamefont{B.}~\bibnamefont{Seeber}}, \bibnamefont{and}
  \bibinfo{author}{\bibfnamefont{C.}~\bibnamefont{Senatore}},
  \bibinfo{journal}{Supercond. Sci. Technol.} \textbf{\bibinfo{volume}{25}},
  \bibinfo{pages}{025015} (\bibinfo{year}{2012}).

\bibitem[{\citenamefont{Lim et~al.}(1983)\citenamefont{Lim, Thompson, and
  Webb}}]{lim83}
\bibinfo{author}{\bibfnamefont{K.~C.} \bibnamefont{Lim}},
  \bibinfo{author}{\bibfnamefont{J.~D.} \bibnamefont{Thompson}},
  \bibnamefont{and} \bibinfo{author}{\bibfnamefont{G.~W.} \bibnamefont{Webb}},
  \bibinfo{journal}{Phys. Rev. B} \textbf{\bibinfo{volume}{27}},
  \bibinfo{pages}{2781} (\bibinfo{year}{1983}).

\bibitem[{\citenamefont{Markiewicz}(2004{\natexlab{b}})}]{markiewicz04bis}
\bibinfo{author}{\bibfnamefont{W.~D.} \bibnamefont{Markiewicz}},
  \bibinfo{journal}{Cryogenics} \textbf{\bibinfo{volume}{44}},
  \bibinfo{pages}{895} (\bibinfo{year}{2004}{\natexlab{b}}).

\bibitem[{\citenamefont{Weber and Goldberg}()}]{weger74}
\bibinfo{author}{\bibfnamefont{W.}~\bibnamefont{Weber}} \bibnamefont{and}
  \bibinfo{author}{\bibfnamefont{I.}~\bibnamefont{Goldberg}} (????).

\bibitem[{\citenamefont{Klein et~al.}(1978{\natexlab{b}})\citenamefont{Klein,
  Boyer, Papaconstantopoulos, and Mattheiss}}]{klein79bis}
\bibinfo{author}{\bibfnamefont{B.~M.} \bibnamefont{Klein}},
  \bibinfo{author}{\bibfnamefont{L.~L.} \bibnamefont{Boyer}},
  \bibinfo{author}{\bibfnamefont{D.~A.} \bibnamefont{Papaconstantopoulos}},
  \bibnamefont{and} \bibinfo{author}{\bibfnamefont{L.~F.}
  \bibnamefont{Mattheiss}}, \bibinfo{journal}{Phys. Rev. B}
  \textbf{\bibinfo{volume}{18}}, \bibinfo{pages}{6411}
  (\bibinfo{year}{1978}{\natexlab{b}}).

\bibitem[{\citenamefont{Mentink et~al.}(2012)\citenamefont{Mentink, Dhalle,
  Dietderich, Godeke, Goldacker, Hellman, and ten Kate}}]{mentink12}
\bibinfo{author}{\bibfnamefont{M.~G.~T.} \bibnamefont{Mentink}},
  \bibinfo{author}{\bibfnamefont{M.~M.~J.} \bibnamefont{Dhalle}},
  \bibinfo{author}{\bibfnamefont{D.~R.} \bibnamefont{Dietderich}},
  \bibinfo{author}{\bibfnamefont{A.}~\bibnamefont{Godeke}},
  \bibinfo{author}{\bibfnamefont{W.}~\bibnamefont{Goldacker}},
  \bibinfo{author}{\bibfnamefont{F.}~\bibnamefont{Hellman}}, \bibnamefont{and}
  \bibinfo{author}{\bibfnamefont{H.~H.~J.} \bibnamefont{ten Kate}},
  \bibinfo{journal}{AIP Conference Proceedings}
  \textbf{\bibinfo{volume}{1435}}, \bibinfo{pages}{225} (\bibinfo{year}{2012}).

\bibitem[{\citenamefont{Welch}(1980)}]{welch80}
\bibinfo{author}{\bibfnamefont{D.~O.} \bibnamefont{Welch}},
  \bibinfo{journal}{Adv. Cryo. Eng.} \textbf{\bibinfo{volume}{26}},
  \bibinfo{pages}{48} (\bibinfo{year}{1980}).

\bibitem[{\citenamefont{Ekin}(1977)}]{ekin77}
\bibinfo{author}{\bibfnamefont{J.}~\bibnamefont{Ekin}}, \bibinfo{journal}{IEEE
  Trans. Magn.} \textbf{\bibinfo{volume}{13}}, \bibinfo{pages}{127}
  (\bibinfo{year}{1977}).

\bibitem[{\citenamefont{Senatore et~al.}(2007)\citenamefont{Senatore,
  Ab\"acherli, Cantoni, and Fl\"ukiger}}]{senatore07}
\bibinfo{author}{\bibfnamefont{C.}~\bibnamefont{Senatore}},
  \bibinfo{author}{\bibfnamefont{V.}~\bibnamefont{Ab\"acherli}},
  \bibinfo{author}{\bibfnamefont{M.}~\bibnamefont{Cantoni}}, \bibnamefont{and}
  \bibinfo{author}{\bibfnamefont{R.}~\bibnamefont{Fl\"ukiger}},
  \bibinfo{journal}{Supercond. Sci. Technol.} \textbf{\bibinfo{volume}{20}},
  \bibinfo{pages}{S217} (\bibinfo{year}{2007}).

\bibitem[{\citenamefont{Godeke}(2006)}]{godeke06bis}
\bibinfo{author}{\bibfnamefont{A.}~\bibnamefont{Godeke}},
  \bibinfo{journal}{Supercond. Sci. Technol.} \textbf{\bibinfo{volume}{19}},
  \bibinfo{pages}{R68} (\bibinfo{year}{2006}).

\end{thebibliography}

\end{document}